\documentclass[conference]{IEEEtran}
\IEEEoverridecommandlockouts
\usepackage[dvips]{graphicx}
\usepackage[english]{babel}
\usepackage{algorithm}
\usepackage{algorithmic}
\usepackage{amsfonts}
\usepackage{amsmath,amssymb}
\usepackage{booktabs}
\usepackage[numbers,sort&compress]{natbib}
\usepackage{bm}
\usepackage{enumerate}
\usepackage{amsfonts,amssymb,amsmath}
\usepackage{multirow}
\usepackage{epstopdf}
\usepackage{stfloats}
\usepackage{caption}
\usepackage{url}
\newtheorem{definition}{\bf{Definition}}

\begin{document}

\title{Joint Deployment of Small Cells and Wireless Backhaul Links in Next-Generation Networks \vspace{-15pt}}

\author{\IEEEauthorblockN{Xiangxiang Xu, Walid Saad, \emph{Senior Member, IEEE}, Xiujun Zhang, \emph{Member, IEEE}, Xibin Xu,\\
\emph{Member, IEEE}, Shidong Zhou, \emph{Member, IEEE}} \vspace{-20pt}

\thanks{This work was partially supported by the Major State Basic Research Development of China (973 program) under Grant 2012CB316002, the National Natural Science Foundation of China under Grant 61201192, the  Science Fund for Creative Research Groups of NSFC under Grant 61321061, the National High Technology Research and Development Program of China (863 program) under Grant 2014AA01A703, the National Science and Technology Major Project of the Ministry of Science and Technology of China under Grant 2014ZX03001014-002,  Tsinghua-Qualcomm Joint Research Program, Tsinghua-Intel International S\&T Cooperation Program (ICRI-MNC), the U.S. Office of Naval Research (ONR) under Grant N00014-15-1-2709.

X. Xu, X. Zhang, X. Xu, S. Zhou are with State Key Laboratory on Microwave and Digital Communications, Department of Electronic Engineering,
Tsinghua University, Beijing 100084, China (e-mail: xuxxmail@163.com, zhangxiujun@tsinghua.edu.cn, xuxb@tsinghua.edu.cn, zhousd@mail.tsinghua.edu.cn).

W. Saad is with Wireless@VT, Bradley Department of Electrical and Computer Engineering, Blacksburg, VA, USA (e-mail: walids@vt.edu). He was also International Scholar with the Department of Computer Engineering, Kyung Hee University, South Korea.
}}

\maketitle

\begin{abstract}
In this paper, a novel approach for optimizing the joint deployment of small cell base stations and wireless backhaul links is proposed.
This joint deployment scenario is cast as a multi-objective optimization problem under the constraints of limited backhaul capacity and outage probability. To address the problem, a novel adaptive algorithm that integrates $\epsilon$-method, Lagrangian relaxation and tabu search is proposed to obtain the Pareto optimal solution set. Simulation results show that the proposed algorithm is quite effective  in finding the optimal solutions. The proposed joint deployment model can be used for planning small cell networks.
\end{abstract}
\IEEEpeerreviewmaketitle
\section{Introduction}
Next-generation cellular networks will be based on the dense deployment of low-cost, low-power small cell base stations (SCBSs) and it is of crucial importance to provide flexible, fast and low cost backhaul solutions to pervasive deployed SCBSs to connect them to the core network~\cite{Andrews14,Taori14,Singh14}. Therefore, to deploy the infrastructure of next-generation cellular systems, backhaul links should be considered jointly with small cells (SCs) in order to optimize network performance.

Most existing works related to base station (BS) deployment have focused mainly on scenarios in which high-speed, fiber enabled wired backhaul sites are available everywhere \cite{Richter12,Brevis11,Ghazzai15,Hsieh14,Zhao14}. However, for dense SC networks, recent works \cite{Senza13} have shown that the availability of such high-speed fiber backhaul links might be scarce. In addition, since SCs must be deployed everywhere, operators must seek alternative backhaul solutions as outlined in~\cite{Senza13,Singh14}. In particular, wireless backhaul solutions such as those based on high frequency millimeter wave (mmW) bands are promising solutions due to the availability of a large bandwidth \cite{Taori14,Singh14}. The authors in \cite{Islam14} studied the placement of wireless backhaul nodes for SCBSs given the location of SCBSs and the traffic load. However, SCBSs deployment is not studied.
While previous works have studied interesting deployment problems, to our best knowledge, the problem of joint deployment of SCs with wireless backhaul links has not been addressed yet.

 The main contribution of this paper is to propose a practical SC-oriented network deployment framework that can explicitly factor in the wired backhaul, wireless backhaul, and statistical user distribution. We formulate the problem as a multi-objective integer programming. To address the joint deployment problem, we introduce an adaptive two-level search algorithm to obtain the entire Pareto optimal solution set. The proposed algorithm enables the system to find the optimal solutions and establishes lower bounds on the quality of those solutions. Simulation results show that the proposed algorithm can find the optimal solutions effectively. In addition, the results show that the joint deployment approach yields a coverage performance that is comparable to a wired backhaul deployment case but with a lower cost.

\section{System Model and Problem Formulation}
Hereinafter, we refer to wireless backhaul nodes as backhaul aggregate nodes (BANs). The BANs need to provide wireless backhaul links to SCBSs. Meanwhile, we assume that BANs have access abilities and can cover a certain area. Hence, BANs act as both aggregators and BSs. Here, SCBSs communicate with BANs using millimeter wave (mmW) links via a point to multi-point (P2MP) transmission mode.


\subsection{Problem Formulation}
\vspace{-4pt}
Consider a geographical area $\cal{A}$ in which a number of SCBSs and BANs must be deployed. The candidate site set is given by ${\cal{F}}={{\cal F}_1}\cup {{\cal F}_2}$ where ${\cal F}_1$
contains sites without fiber connections while locations in ${\cal F}_2$
have fiber backhaul connections. Since BANs need to connect the SCBSs to the core network, the potential sites set for BANs is ${\cal F}_2$. For SCBSs, the candidate sites set is ${\cal {F}}_1$.

 We define binary deployment variables $y_{i} \in \{0,1\}$ and $z_{k}\in \{0,1\}$ for SCBSs and BANs at locations $i \in {\cal F}_1$ (SCBS $i$) and $k \in {\cal F}_2$ (BAN $k$) to indicate the facility is deployed $(=1)$ or not $(=0)$.
 In order to cope with a continuous user distribution, area $\cal{A}$ is partitioned equally to a set of subareas which are denoted by ${\cal{S}}=\{1,\ldots,S\}$. Mobile users are assumed to be distributed according to a homogeneous Poisson point process (PPP) of density ${\lambda}_u$ \cite{Singh14}. Binary variables $x_{ij}$, $x_{kj}\in \{0,1\}$ are defined to indicate the coverage of subarea $j$ by SCBS $i$, BAN $k$ and $x_{ki}\in \{0,1\}$ is defined to indicate the availability of a connection link between SCBS $i$ and BAN $k$.

Given area ${\cal A}$, our goal is to find an optimal deployment of SCBSs, BANs and their connections. Two objective functions are considered: deployment cost and area coverage. So our objective function is given by $\bm{f}\left(\bm{x},\bm{y},\bm{z}\right)=\left[{f_1}\left(\bm{y},\bm{z}\right),{f_2}\left({\bm{x}}\right)\right]$ where
\begin{align}
    &{f_1}\left(\bm{y},\bm{z}\right) = \sum\limits_{{i} \in {{\cal F}_1}} {{y_{{i}}}{c_{{i}}}}  + \sum\limits_{{k} \in {{\cal F}_2}} {{z_{{k}}}{c_{{k}}}},\label{F1}
\end{align}
\begin{align}
    &{f_2}\left({\bm{x}}\right) = S-\sum\limits_{j \in {\cal S}} {\sum\limits_{{i} \in {\cal F}_1} {{x_{ij}}} }-\sum\limits_{j \in {\cal S}} {\sum\limits_{{k} \in {\cal F}_2} {{x_{kj}}} }.\label{F2} \vspace{-5pt}
\end{align}
In (\ref{F1}), $c_i$ and $c_k$ denote the deployment cost of SCBS $i$ and BAN $k$. ${\bm{x}}=\left[x_{ij},x_{kj},x_{ki}\right]$, ${\bm{y}}=\left[y_i\right]$ and ${\bm{z}}=\left[z_k\right]$ are the binary variables. ${f_2}\left({\bm{x}}\right)$ represents the number of  uncovered subareas.

For each subarea $j$, we choose its center point to indicate its coverage. A subarea $j$ is said to be \emph{covered} when its outage probability $p_{oa}$ is satisfied by a deployed BS. It is well-known \cite{Ghosh14,Singh14} that large bandwidth mmW networks are noise limited. Hence, the outage probability can be expressed in terms of signal to noise ratio (SNR).
Formally, the deployment problem can be formulated as a multi-objective integer programming problem:
\begin{equation}
\textrm{(P)}\qquad \mathop {\min }\limits_{{\bm{x}},{\bm{y}},{\bm{z}}} \quad \bm{f}\left(\bm{x},{\bm{y}},{\bm{z}}\right),  \label{PObj} \nonumber
\end{equation}
\vspace{-15pt}
\begin{align}
\textrm{s.t.}\quad &{x_{ij}} \le {y_{i}} \qquad j \in {\cal S},i \in {\cal F}_1, \label{OpenToConnectAccess1} \\
&{x_{kj}} \le {z_k} \qquad j \in {\cal S},k \in {\cal F}_2, \label{OpenToConnectAccess2}\\
&x_{ki} \le z_k \qquad i \in {\cal F}_1, k \in {\cal{F}}_2, \label{OpenToConnectBackhaul2}\\
&\sum\limits_{i \in {\cal F}_1} {x_{ij}} +   \sum\limits_{k \in {\cal F}_2} {x_{kj}}\leq 1 \qquad j \in {\cal S}, \label{AtMostOneAccess}\\
&x_{ij}p({{\gamma}}_{ij}\le \gamma) + x_{kj}p({{\gamma}}_{kj}\le \gamma) \le p_{oa}, \label{SNRAccess} \\
&\sum\limits_{i \in {{\cal F}_1}} {{x_{ki}}}  \le {N_b} \qquad {k} \in {{\cal F}_2}, \label{OpenToConnectBackhaul} \\
&\sum\limits_{k \in {\cal F}_2} {x_{ki}}  = {y_i} \qquad {i} \in {{\cal F}_1}, \label{OpenMustServedBackhaul}\\
&x_{ki}p(r_{i,total} \ge C_{ki}) \le p_{bb} \qquad i \in {{\cal F}_1}, k \in {{\cal F}_2}, \label{BlockBackhaul}
\end{align}

Here, ${{\gamma}}_{ij}$ (${{\gamma}}_{kj}$) is the received SNR at the center point of area $j$ from SCBS $i$ (BAN $k$) and $\gamma$ is the SNR threshold. $r_{i,total}$ is defined as the sum of requested user rates from the covered area of SCBS $i$. $C_{ki}$ denotes the capacity of the link from BAN $k$ to SCBS $i$. $p_{bb}$ is the maximum allowable probability with which the limited backhaul capacity introduces a large delay for users.

In problem (P), constraints (\ref{OpenToConnectAccess1}), (\ref{OpenToConnectAccess2}), (\ref{AtMostOneAccess}) and (\ref{SNRAccess}) ensure that each subarea can be covered at most by one \emph{deployed} BS with lower outage probability than the threshold. (\ref{OpenToConnectBackhaul}) means that at most $N_b$ small cells can be served by one BAN simultaneously. (\ref{OpenMustServedBackhaul}) indicates that wireless backhaul is needed once an SCBS is deployed.
Constraints (\ref{OpenToConnectBackhaul2}) and  (\ref{OpenMustServedBackhaul}) together ensure that a deployed SCBS must be served by a deployed
BAN. (\ref{BlockBackhaul}) characterizes the wireless backhaul capacity constraints. Here, we mainly focus on limited backhaul capacity and assume that access capacity is sufficient for users.

Consider SCBS $i$, the mean number of users in its coverage is ${\lambda}_i={{\lambda}_u}{\Delta s}\sum\limits_{j \in {\cal S}}{x_{ij}}$. Then the probability in (\ref{BlockBackhaul}) is
\vspace{-10pt}
\begin{align}
 p({r_{i,total}} \ge {C_{ki}}) ={e^{ - {\lambda _i}}}\sum\limits_{q = 1}^\infty  {p(\sum\limits_{m = 1}^q {{r_m}}  \ge {C_{ki}})\frac{{\lambda _i^q}}{{q!}}}, \label{DetailedCapacityCon} \vspace{-5pt}
\end{align}
where $r_m$ is the random rate demand of the $m$ th user.
Given the users' rate demand distribution, from (\ref{BlockBackhaul}) and (\ref{DetailedCapacityCon}), we can then derive the following constraints on coverage for SCBS $i$:
\begin{align}
\sum\limits_{j \in {\cal S}}{x_{ij}}\le \sum\limits_{k \in {\cal F}_2} {N_{ki}x_{ki}}   \quad i \in {\cal F}_1, \label{NumberLimit}
\vspace{-10pt}
\end{align}
where $N_{ki}$ is the maximum number of subareas that SCBS $i$ can cover when it is connected to BAN $k$.

Problem (P) can be cast within the framework of the \emph{two-level capacitated facility location problem and multi-objective combinatorial optimization} \cite{Hamacher02}. To solve such a problem, the notion of Pareto optimality is commonly used \cite{Laumanns06}:
\begin{definition}
  An objective vector ${\bm{f}}(\bm{x}^*)=\left[f_1^*,\dots,f_N^*\right]$ is {\emph{Pareto optimal}} or {\emph{nondominated}} if and only if there does not exist another objective vector $\bm{f}(\bm{x})$ such that $f_i \le f_i^*, \forall i$ and $f_i < f_i^*, \exists i$. $\bm{x}^*$ is {\emph{efficient}} if ${\bm{f}}(\bm{x}^*)$ is Pareto optimal. $\bm{x}^*$ is called {\emph{weakly efficient}} if there is no other $\bm{x}$ such that $f_i < f_i^*, \forall i$.
\end{definition}

 Hereinafter, we use Pareto optimal and efficient interchangeably.
 Our goal is to find the Pareto optimal solution set for decision makers to guide the practical deployment.
\vspace{-5pt}
 \subsection{Millimeter-wave band and propagation model}
\vspace{-5pt}
%
For mmW propagation, we use the path loss model proposed in \cite{Ghosh14} for the urban scenarios:
\begin{align}
L(d)=20{\log _{10}}\left(\frac{{4\pi {d_0}}}{\lambda }\right){\rm{ + }}10\bar n{\log _{10}}\left(\frac{d}{{{d_0}}}\right) + \chi, \label{PropagationModel}
\end{align}
where $d_0$ is a reference distance and $\lambda$ is the wavelength of the carrier. $\bar n$ is the path loss exponent and $d$ is the transmission distance in meters. $\chi$ is the shadowing component which is a zero mean Gaussian variable with a standard deviation ${\sigma}_s$ in dB.
The model
in \cite{Kulkarni14} is utilized in our work which assumes the LOS probability as a function of transmission distance, $p_{LOS}\left(d\right)=e^{-{\beta}d}$, where $\beta$ is related with transmission environment.
Based on this model, constraint (\ref{SNRAccess}) can be transformed further to a constraint on transmission distance $d_{ij}$ and $d_{kj}$ as follows:
\begin{align}
x_{ij}\left(d_{ij}-D_{it}\right)+x_{kj}\left(d_{kj}-D_{kt}\right) \le 0. \label{DistanceConstraint}
\end{align}
And $d_{ij} \left(d_{kj}\right)$ is defined as the distance from SCBS $i$ (BAN $k$) to the center point of subarea $j$.

\vspace{-6pt}
\section{Proposed Algorithm}
\vspace{-4pt}
One approach to solve (P) is to adopt a weighted sum method.
However, such an approach can only find supported solutions. In problem (P), it is not guaranteed that all Pareto optimal solutions are supported due to the binary variables.

The $\epsilon$-constraint method is another approach in multi-objective programming that can enumerate the entire Pareto solutions set. 
All but one objective are taken into constraints. Through the update of the parameter $\epsilon$, the entire efficient solution set can be obtained \cite{Laumanns06}. Here, considering the combinatorial structure, we propose an adaptive $\epsilon$-constraint method to find the entire Pareto optimal solution set. Coverage is maximized with constrained deployment cost. Then we can get the following problem:
\begin{equation}
\left(\textrm{P}_{\epsilon}\right)\quad \mathop {\min } \quad S-\sum\limits_{j \in {\cal S}} {\sum\limits_{{i} \in {\cal F}_1} {{x_{ij}}} }-\sum\limits_{j \in {\cal S}} {\sum\limits_{{k} \in {\cal F}_2} {{x_{kj}}} }, \nonumber
\end{equation}
\vspace{-12pt}
\begin{align}
\textrm{s.t.} \quad &\sum\limits_{{i} \in {{\cal F}_1}} {{y_{{i}}}{c_{{i}}}}  + \sum\limits_{{k} \in {{\cal F}_2}} {{z_{{k}}}{c_{{k}}}} \le {\epsilon}, \label{CostConstraints} \\
&\textrm{constraints } (\ref{OpenToConnectAccess1}-\ref{AtMostOneAccess}),(\ref{OpenToConnectBackhaul}),(\ref{OpenMustServedBackhaul}),(\ref{NumberLimit}),(\ref{DistanceConstraint}). \nonumber
\end{align}
\subsection{Lagrangian Relaxation}
\vspace{-3pt}
Here, by introducing Lagrangian multipliers and relaxing constraints  (\ref{OpenToConnectBackhaul}) and  (\ref{NumberLimit}), we get a relaxed problem ($\textrm{P}_{\bm{\lambda}}$).
\begin{equation}
\left(\textrm{P}_{\bm{\lambda}}\right)\quad \mathop {\min } \quad S - \sum\limits_{i \in {{\cal F}_1}} {{m_i}{y_i}}  - \sum\limits_{k \in {{\cal F}_2}} {{n_k}{z_k}}-\sum\limits_{k \in {\cal F}_2}{{\lambda}_{1k}N_b}
 \nonumber
\end{equation}
\begin{align}
\textrm{s.t.} \quad \textrm{constraints } (\ref{OpenToConnectBackhaul2}),(\ref{AtMostOneAccess}), (\ref{OpenMustServedBackhaul}), (\ref{DistanceConstraint}) \textrm{ and } (\ref{CostConstraints}).\nonumber
\end{align}
where
$
m_i=\sum\limits_{j \in {\cal S}}{\left(1-{\lambda}_{2i}\right)x_{ij}}-\sum\limits_{k \in {\cal F}_2}{\left({\lambda}_{1k}-{\lambda}_{2i}N_{ki}\right)x_{ki}}$ and
$n_k=\sum\limits_{j \in {\cal S}}{x_{kj}}$.
${\lambda}_{1k},{\lambda}_{2i}$ are the nonnegative Lagrangian variables for  constraints (\ref{OpenToConnectBackhaul}) and  (\ref{NumberLimit}), respectively. ($\textrm{P}_{\bm{\lambda}}$) is not simpler than the knapsack problem since the different items here, SCs and BANs, are correlated with each other through variables $x_{ki}$, $x_{ij}$ and $x_{kj}$.
However, if we relax  (\ref{OpenToConnectBackhaul2}) or (\ref{AtMostOneAccess}) further, the obtained optimal value of ($\textrm{P}_{\bm{\lambda}}$) would degrade significantly.

Since ($\textrm{P}_{\bm{\lambda}}$) is a relaxation of minimization problem ($\textrm{P}_{\epsilon}$), a lower bound of ($\textrm{P}_{\epsilon}$), $B_{lb}$, can be obtained through solving ($\textrm{P}_{\bm{\lambda}}$). And this lower bound can be improved iteratively through the update of $\bm{\lambda}$. A subgradient method is utilized to update $\bm{\lambda}$.

\subsection{Tabu Search}

To solve ($\textrm{P}_{\bm{\lambda}}$) first, we propose a tabu search based algorithm, described in Algorithm 1, to find the optimal solution. Here, open, close and swap moves are defined to search the neighborhood, ${\cal N}([\bm{y},\bm{z}])$, of a feasible solution $[\bm{y},\bm{z}]$. Restart diversification scheme is defined when all candidate solutions are tabu solutions.


\begin{table}[t]\scriptsize
\begin{tabular}
{p{0.95\columnwidth}}
\\
\toprule
\textbf{Algorithm 1} Tabu search for problem (${\textrm{P}}_{\bm{\lambda}}$).\\
\midrule
\textbf{Input:} Initial solution $\bm{q}=\left[{\bm{y}},{\bm{z}}\right]$, $N_{max}$, ${\lambda}_{1k}$, ${\lambda}_{2i}$, $c_i$, $c_k$, $T_{div}$, $N_{div}$, $N_b$.
\begin{enumerate}[  1:]
\item initialize ${\bm{q}}_0=\bm{q}$, $t=0$, $\bm{q}_B=\bm{q}_0$, objective value $V({\bm{q}_B})$, empty tabu list
\item \textbf{while ($t<N_{max}$)}
\end{enumerate}
\begin{enumerate}[  1:\quad]
\setcounter{enumi}{2}
\item compute ${\cal N}({{\bm{q}}_{t}})$, denote $\bm{q}_b=\mathop {\arg \min }\limits_{\bm{q} \in {{\cal N}({\bm{q}}_{t})}}{V({\bm{q}})}$
\item aspiration criterion: if $V(\bm{q}_b)<V({\bm{q}}_B)$, $\bm{q}_B=\bm{q}_b$ and $\bm{q}_n=\bm{q}_b$
\item if $V(\bm{q}_b)\ge V({\bm{q}}_B)$, denote $\bm{q}_n=\mathop {\arg \min }\limits_{\bm{q} \in {{\cal N}({\bm{q}}_{t})}, \textrm{ }\bm{q} \textrm{ is non-tabu}}{V({\bm{q}})}$
\item diversification: if ${\bm{q}}_B$ does not change for $T_{div}$ steps or no non-tabu solutions exist, deploy $N_{div}$ rarely deployed BSs and denote the new solution as ${\bm{q}}_n$
\item $t=t+1$, $\bm{q}_{t}=\bm{q}_n$, update tabu list
\end{enumerate}
\begin{enumerate}[  1:]
\setcounter{enumi}{7}
\item \textbf{end while}
\end{enumerate}
\textbf{Output:} optimized solution $\bm{q}_B=\left[{\bm{y}}^*,{\bm{z}}^*\right]$ and optimized objective value for (${\textrm{P}}_{\bm{\lambda}}$). \\
\bottomrule
\end{tabular}
\vspace{-18pt}
\end{table}

Note that the solution provided by Lagrangian can be infeasible to ($\textrm{P}_{\epsilon}$) due to the relaxation. We denote the solution obtained from Lagrangian relaxation as $\left[{\bm{x}},{\bm{y}},\bm{z}\right]$.
Modification procedures are defined here to modify relaxed solutions.
\vspace{-3pt}
 \begin{enumerate}[Step 1:]
   \item Check the (\ref{OpenToConnectBackhaul}) from the perspective of total number. If the deployed BANs can not serve deployed SCs, SCs are removed starting from the one with minimum $m_i$ until the constraint is satisfied. Reconnect SCBSs with BANs if (\ref{OpenToConnectBackhaul}) is violated for some BANs.
   \item Check the constraint (\ref{NumberLimit}). For SCs with more coverage than backhaul constraint, covered area is reduced from the cell edge until the constraint is satisfied.
   \item Check the available connection for subareas not covered. For these subareas with available connections, the base station with the minimum ${\lambda}_i$ is chosen.
 \end{enumerate}
\vspace{-3pt}
After the modification, the feasible solution to problem ($\textrm{P}_{\epsilon}$) can be obtained and it is denoted by $\left[{\bm{x}}',{\bm{y}}',\bm{z}'\right]$. However, this solution may not be efficient due to the modification process. Here, we propose a two-level tabu search algorithm, Algorithm 2, for multi-objective programming to iteratively improve feasible solutions and get a set of approximate non-dominated solutions. Furthermore, weakly efficient solutions obtained by $\epsilon$ method can be eliminated.
The two-level structure is also a diversification scheme which enforces the sufficient search of BAN space. Here, $\Delta\epsilon$ is defined to control the intensification by the decision maker.

\begin{table}[t]\scriptsize
\begin{tabular}
{p{0.95\columnwidth}}
\\
\toprule
\textbf{Algorithm 2} Two-level tabu search algorithm.\\
\midrule
\textbf{Input:} feasible solution $\bm{p}=\left[{\bm{x}},{\bm{y}},\bm{z}\right]$, $N_{t_1},N_{t_2}$, $N_{swap}$, $N_{div}$, $\epsilon$, $\Delta\epsilon$.
\begin{enumerate}[  1:]
\item initialize ${\bm{p}}_0=\bm{p}$, $t_1=0,t_2=0$, ${\cal{P}}_e=\bm{p}$, empty tabu list
\item \textbf{while ($t_1<N_{t_1}$)}
\end{enumerate}
\begin{enumerate}[  1:\quad]
\setcounter{enumi}{2}
\item compute ${\cal N}_{\bm{z}}\left({\bm{p}}_{t_1}\right)$ for BANs with fixed $\bm{y}$, ${\cal P}_t=\{\bm{p}|\bm{p} \in {{\cal N}_{\bm z}({\bm p}_{t_1})}, f_1(\bm p) \in [\epsilon-\Delta\epsilon,\epsilon], \bm{p}$ is nondominated\}.
\item ${\cal{P}}_e={\cal{P}}_e\cup {{\cal{P}}_t}$, delete dominated solutions in ${\cal{P}}_e$ \label{Itera2}
\item denote ${\bm{p}}_{t_2} = \mathop {\arg \min }\limits_{{\bm{p}} \in {{\cal{P}}_t}, {\bm{p}} \textrm{ is non-tabu} }{f_2(\bm{p})}$, $t_1=t_1+1$, update tabu list
\item \textbf{while ($t_2<N_{t_2}$)}
\end{enumerate}
\begin{enumerate}[  1:\qquad]
\setcounter{enumi}{6}
\item compute ${\cal N}_{\bm{y}}\left({\bm{p}}_{t_2}\right)$ for SCs with fixed $\bm z$, ${\cal P}_t=\{\bm{p}|\bm{p} \in {{\cal N}_{\bm y}({\bm p}_{t_2})}, f_1(\bm p) \in [\epsilon-\Delta\epsilon,\epsilon], \textrm{ } \bm{p}$ is nondominated\}.
\item  ${\cal{P}}_e={\cal{P}}_e\cup {{\cal{P}}_t}$, delete dominated solutions in ${\cal{P}}_e$
\item $t_2=t_2+1$, denote ${\bm{p}}_{t_2} = \mathop {\arg \min }\limits_{{\bm{p}} \in {{\cal{P}}_t}, {\bm{p}} \textrm{ is non-tabu} }{f_2(\bm{p})}$, update tabu list
\end{enumerate}
\begin{enumerate}[  1:\quad]
\setcounter{enumi}{9}
\item \textbf{end while}
\item $\bm{p}_{t_1}=\bm{p}_{t_2}$ and $t_2=0$
\end{enumerate}
\begin{enumerate}[  1:]
\setcounter{enumi}{11}
\item \textbf{end while}
\end{enumerate}
\textbf{Output:} Nondominated solution set ${\cal{P}}_e$. \\
\bottomrule
\end{tabular}
\vspace{-16pt}
\end{table}
\subsection{Update of $\epsilon$}
In order to get the exact Pareto optimal frontier, $\epsilon$ needs to be decreased gradually. Here, we use an adaptive update scheme. In iteration $t$, suppose the cost limit is ${\epsilon}_t$. Through solving problem $\textrm{P}_{{\epsilon}_t}$, we can get optimal solution $\left[{\bm{x}}_t, {\bm{y}}_t, {\bm{z}}_t\right]$ and objective function $f_1^t, f_2^t$. The cost constraint is then updated as
${\epsilon}_{t+1}={f_1^t}-\Delta c$. $\Delta c$ is a small positive number.
 Since we can get multiple efficient solutions through algorithm 2 in one iteration, a more general update equation is given as
\vspace{-5pt}
\[
{\epsilon}_{t+1}={\min}\left(\mathop {\min }\limits_{\left[{\bm{y}}_t,{\bm{z}}_t\right] \in {{\cal{P}}_e}} {f_1}({\bm{y}}_t,{\bm{z}}_t), {\epsilon}_t\right) -\Delta c.
\vspace{-3pt}
\]
The algorithm to solve (P) 
 is summarized as Algorithm 3.

\begin{table}[t]\scriptsize
\begin{tabular}
{p{0.95\columnwidth}}
\\
\toprule
\textbf{Algorithm 3} Algorithm to solve (P).\\
\midrule
\textbf{Input:} locations of potential sites, $c_i$, $c_k$, $N_b$, $N_{max,L}$, $\Delta\epsilon$.
\begin{enumerate}[  Step 1:]
\item initialize ${\epsilon}_0=\sum\limits_{i \in {\cal{F}}_1}{c_i}+\sum\limits_{k \in {\cal{F}}_2}{c_k}$, $\cal{P}=\emptyset$, $t=0$, $t_L=0$
\item solve (${\textrm{P}}_{\bm{\lambda}}$) with ${\epsilon}_t$, get solution $\bm{q}=\left[{\bm{y}},{\bm{z}}\right]$ \label{SolveLagrangian}
\item modify solution $\bm{q}$ and get a feasible solution $\bm{q}'$
\item two-level tabu search from $\bm{q}'$ and get nondominated solution set ${\cal{P}}_e$
\item ${\cal{P}}={\cal{P}} \cup {{\cal{P}}_e}$, update $\bm{\lambda}$ through the solution in ${\cal{P}}_e$ with maximum cost,  $t_L=t_L+1$
\item if $t_L<N_{max,L}$, return to step \ref{SolveLagrangian}
\item update $\epsilon$, $t=t+1$
\item if ${\epsilon}_t>0$, return to step \ref{SolveLagrangian}
\end{enumerate}
\textbf{Output:} $\cal{P}$, a cost constraint set $\cal{C}$ and a lower bound set ${\cal{B}}_{lb}$. \\
\bottomrule
\end{tabular}
\vspace{-7pt}
\end{table}
\section{Simulation Results And Discussions}
\vspace{-5pt}
In this section, simulation results are given to evaluate the proposed deployment algorithm. In our simulations, a 73GHz mmW is considered. Channel model parameters in (\ref{PropagationModel}) are shown in Table \ref{Channel parameters} with $d_0=1 \textrm{m}$ \cite{Ghosh14}.

\begin{table}[t]\scriptsize
\setlength{\abovecaptionskip}{0pt}
\setlength{\belowcaptionskip}{0pt}
\centering
\caption{\small Channel model parameters} \label{Channel parameters}
\begin{tabular}{|c|c|c|c|c|c|}
  \hline
    &\multicolumn{2}{c|}{Backhaul link} &\multicolumn{2}{c|}{Access link} &\multirow{2}{*}{$\beta$}\\
  \cline{1-5}
    &$\bar n$ &${\sigma}_s$(dB) &$\bar n$ &${\sigma}_s$(dB)  &\\
    \hline
  LOS &2.0  &4.2 &2.0 &5.2  &\multirow{2}{*}{0.046}\\
  \cline{1-5}
  NLOS &3.5  &7.9 &3.3 &7.6  &\\
  \hline
 \end{tabular}
 \vspace{-8pt}
\end{table}
\begin{table}[t]\scriptsize
\setlength{\abovecaptionskip}{0pt}
\setlength{\belowcaptionskip}{0pt}
\centering
\caption{\small Simulation parameters} \label{Simulation parameters}
\begin{tabular}{ccc}
\toprule
 Symbol &Meaning &Value   \\
\midrule
${\lambda}_u$  &Intensity of users &200/${\textrm{km}}^2$ \\
$P_t$  &Transmit power &30dBm  \\
$c_i$   &Normalized deployment cost of an SCBS &1\\
$c_k$  &Normalized deployemnt cost of a BAN &10  \\
$N_b$ &Maximum number of SCs served by one BAN &3\\
 \bottomrule
 \end{tabular}
 \vspace{-12pt}
\end{table}
Table \ref{Simulation parameters} shows the simulation parameters. For users, the SNR threshold is -10 dB which was calculated from a 100Mbps outage rate with a 1GHz bandwidth \cite{Ghosh14}. The noise power is ${\sigma}^2$  is $-74$dBm \cite{Kulkarni14}.
The subarea is $10\textrm{m}\times10\textrm{m}$ in our simulation. For simulations, we tailor the conventional tabu search for our problem and compare its performance with the proposed algorithm which are hereinafter referred to as `Single-level Tabu` and `Proposed`,  respectively.

\begin{figure}[!t]
\setlength{\abovecaptionskip}{10pt}
\setlength{\belowcaptionskip}{0pt}
\centering
\scalebox{0.34}{\includegraphics*[11,5][492,375]{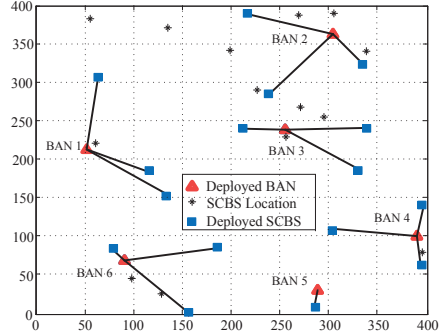}}
\vspace{-7pt}
\caption{Deployment instance when the number of available sites is 36.}
\label{Instance}
\vspace{-13pt}
\end{figure}

Fig. \ref{Instance} shows a deployment instance. Here, 6 available sites for BANs are all utilized and 16 SCBSs are placed. It is indicated from the results that improved coverage performance can be obtained through the utilization of sites without fiber connections.
Also, due to the noise limited nature of mmW band, some SCBSs are placed close to one another.

\begin{figure}[!t]
\setlength{\abovecaptionskip}{10pt}
\setlength{\belowcaptionskip}{0pt}
\centering
\scalebox{0.34}{\includegraphics*[6,5][506,392]{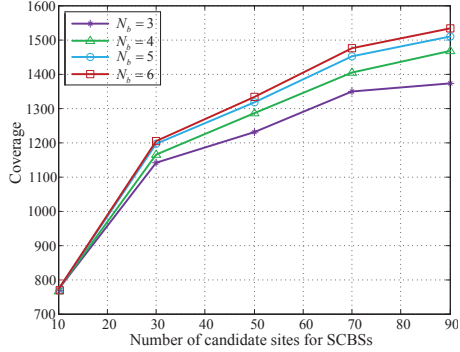}}
\vspace{-7pt}
\caption{Coverage results for different number of candidate sites for SCBSs.}
\label{PerformanceComparison}
\vspace{-15pt}
\end{figure}

In Fig. \ref{PerformanceComparison}, we show how the coverage performance varies with the increase in candidate sites for SCBSs given 6 potential sites for BANs in an $400~\textrm{m}\times400~\textrm{m}$ area. With the increase of candidate sites, the coverage performance can be improved significantly. Also, performance can be enhanced furthermore by increasing the capabilities of BANs ($N_b$). When the number of SCBS's candidate sites is low, increasing the candidate sites can bring more benefit than improving capability of a BAN. So instead of increasing the capability of a BAN, it is recommended to find a large number of candidate sites to guarantee a good coverage. When the amount of available sites is large enough (70 in Fig. \ref{PerformanceComparison}), increasing the capabilities of BANs gains more coverage and this gain decreases with the increase of $N_b$.

\begin{figure}[!t]
\setlength{\abovecaptionskip}{10pt}
\setlength{\belowcaptionskip}{0pt}
\centering
\scalebox{0.38}{\includegraphics*[1,1][494,392]{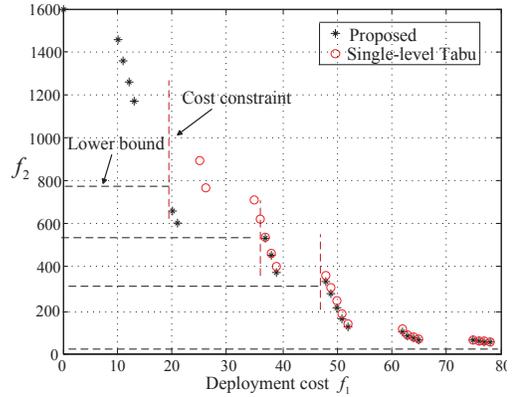}}
\vspace{-7pt}
\caption{Pareto optimal solutions for an $400\textrm{m}\times400\textrm{m}$ area with 6 available sites for BANs and 30 available sites for SCBSs.}
\label{Results400}
\vspace{-16pt}
\end{figure}

In Fig. \ref{Results400}, we compare the obtained Pareto optimal solutions for an $400\textrm{m}\times400\textrm{m}$ area. Here, the proposed algorithm obtains better optimal solutions due to the good initial solution provided by Lagrangian relaxation and the two-level diversification scheme compared with single-level tabu search. Fig. \ref{Results400} also shows that the coverage performance of 4 BANs with 9 SCBSs is almost the same with that of 5 BANs which is an example of wired backhaul network.
 This indicates that, besides flexibility, a comparable  coverage performance can be obtained with wireless backhaul SCBSs which leads to a significant cost reduction in practical scenario where rental cost is much higher than parameter set.

\section{Conclusion}
In this paper, we have studied the joint deployment of SCs and wireless backhaul links in next-generation networks. A general two objective integer optimization model has been proposed for optimizing the location of SCBSs, BANs and their connections. The proposed model jointly considers the outage probability and backhaul link capacity constraints. To obtain the Pareto optimal solution set, we have proposed a two-level adaptive algorithm based on  $\epsilon$-method, Lagrangian relaxation and tabu search. Simulation results have shown that our algorithm can obtain the optimal solution set effectively.


\small
\setlength{\baselineskip}{0.95\baselineskip}
\def\baselinestretch{0.78}
\footnotesize
\bibliographystyle{IEEEtran}
\bibliography{refs_paper}

\end{document}